# Machine Learning-Assisted Surrogate Modeling with Multi-Objective Optimization and Decision-Making of a Steam Methane Reforming Reactor

Seyed Reza Nabavi [1,*], Zonglin Guo [2], Zhiyuan Wang [3]

1. Department of Applied Chemistry, Faculty of Chemistry, University of Mazandaran, Babolsar, 47416-95447, Iran

2. School of Electrical and Electronic Engineering, Shanghai Institute of Technology, Shanghai 201418, China

3. School of Business, Singapore University of Social Sciences, Singapore 599494, Singapore

* Corresponding author: Seyed Reza Nabavi, Email: srnabavi@umz.ac.ir

## Abstract

This study presents an integrated modeling and optimization framework for a steam methane reforming (SMR) reactor, combining a mathematical model, artificial neural network (ANN)-based hybrid modeling, advanced multi-objective optimization (MOO) and multi-criteria decision-making (MCDM) techniques. A one-dimensional fixed-bed reactor model accounting for internal mass transfer resistance was employed to simulate reactor performance. To reduce the high computational cost of the mathematical model, a hybrid ANN surrogate was constructed, achieving a 93.8% reduction in average simulation time while maintaining high predictive accuracy. The hybrid model was then embedded into three MOO scenarios using the non-dominated sorting genetic algorithm II (NSGA-II) solver: 1) maximizing methane conversion and hydrogen output; 2) maximizing hydrogen output while minimizing carbon dioxide emissions; and 3) a combined three-objective case. The optimal trade-off solutions were further ranked and selected using two MCDM methods: technique for order of preference by similarity to ideal solution (TOPSIS) and simplified preference ranking on the basis of ideal-average distance (sPROBID). Optimal results include a methane conversion of 0.863 with 4.556 mol/s hydrogen output in the first case, and 0.988 methane conversion with 3.335 mol/s hydrogen and 0.781 mol/s carbon dioxide in the third. This comprehensive methodology offers a scalable and effective strategy for optimizing complex catalytic reactor systems with multiple, often conflicting, objectives.

## 1. Introduction

In the 20th century, with the growth of industrial processes and chemical industry technology, the demand for crude oil increased in such a way that it took a significant share in the supply of energy and raw materials (Lang & Auer, 2020). In recent decades, the unprecedented growth of the population, accompanied by rising consumerism, has raised concerns about the depletion of oil resources. As a result, measures have been taken to control and manage consumption, and most countries are now seeking ways to use oil more efficiently. Studies have showed that natural gas and coal have a good potential to replace crude oil in the future. The reason for this claim is the key intermediate production of synthesis gas from natural gas and coal. Synthesis gas is a mixture of hydrogen and carbon monoxide with different proportions, which can be the beginning of the production chain of petrochemical products as well as fuel. In addition, from the perspective of renewable sources, synthesis gas can be produced through biomass gasification or other processes associated with methane generation (Wang et al., 2022; Siang et al., 2024). One of the most basic industrial methods of synthesis gas production is steam methane reforming (SMR) process. The popularity of this process can be attributed to its high efficiency compared to other processes (Abdulrasheed et al., 2019). Ben-Mansour et al. (2023) performed a computational study on a membrane-integrated SMR reactor to evaluate how different thermal boundary conditions, such as constant and variable wall temperatures and heat fluxes, affect methane conversion and hydrogen recovery. Dong et al. (2024) experimentally investigated photo-thermal SMR process over $Co/Al_2O_3$ catalysts and demonstrated that light irradiation could enhance hydrogen production. Song et al. (2025) proposed and analyzed a sorption-enhanced electrified SMR process that integrates calcium looping and renewable electricity to improve thermal efficiency, enable in-situ $CO_2$ capture, and achieve near-zero-emission hydrogen production.

SMR process modeling is essential for process control and optimization. The effects of various process variables, such as inlet temperature, water-to-methane feed ratio, and hydrogen-to-methane feed ratio, on methane conversion, as well as hydrogen and syngas productivity, can be systematically investigated. In the present research, the first stage involves mathematical modeling of SMR process in a fixed-bed reactor using a one-dimensional model that accounts for the internal mass transfer resistance within the catalyst particles (Figure S1). Subsequently, temperature profiles, conversion rates, and species mole fractions are calculated along the length of the reactor. In the SMR process, reforming reactions occur within the interior of the catalyst particles and are often governed by internal mass transfer limitations. The one-dimensional descriptive model of this phenomenon involves solving a system of boundary value differential equations numerically. Furthermore, integrating the conservation equations with this system significantly increases the computational time required for solving the equations, which limits the practical applicability of the mathematical model in process control and optimization tasks. Hybrid modeling has been proposed as a solution to address these challenges(Parisi & Laborde, 2001). In this work, a hybrid model is developed to accelerate the computation of species profiles in the SMR reactor. Neural networks models (Goodfellow et al., 2016) are utilized in constructing the hybrid model for the SMR reactor. This approach is expected to significantly increase computational speed and enable faster prediction of species profiles. Next, the developed hybrid model is incorporated when performing multi-objective optimization (MOO) of the SMR process considering various objective functions.

Reviews of MOO applications in chemical engineering highlight the prevalence of the elitist non-dominated sorting genetic algorithm II (NSGA-II). NSGA-II was originally proposed by Deb et al. (2002). For instance, Shahhosseini et al. (2016) implemented NSGA-II to design an industrial membrane SMR reactor optimized for syngas production tailored to Fischer–Tropsch synthesis, balancing $CH_4$ conversion, $H_2$ selectivity, and CO selectivity. Nabavi (2016) applied NSGA-II to simultaneously maximize membrane flux and separation factor in the pervaporation of isopropanol–water mixtures, by optimizing the preparation conditions of asymmetric polyetherimide membranes, including polymer concentration, additive content, solvent evaporation temperature, and time. Chen et al. (2022) used NSGA-II for MOO of a SMR membrane reactor heated by molten salt, achieving higher hydrogen production and lower entropy generation. Nabavi et al. (2024) utilized NSGA-II in designing a multi-tubular packed-bed membrane reactor for the oxidative dehydrogenation of ethane, maximizing ethane conversion and ethylene selectivity, and minimizing carbon dioxide emissions. Gao et al. (2024) developed a Kriging-assisted NSGA-II model to identify Pareto-optimal designs for methane-fueled slit-type combustors, optimizing radiation efficiency, standard deviation, and volume power density based on varying wall thermal conductivities. Dai et al. (2025) developed a data-driven NSGA-II approach to optimize geometric parameters of a methane pre-reformer for solid oxide fuel cell systems, maximizing methane conversion and minimizing costs simultaneously. Wang et al. (2025) successfully integrated NSGA-II into reinforcement learning framework for nonlinear chemical processes. These studies underscore the versatility and robustness of NSGA-II in handling the complex trade-offs inherent in chemical process design and optimization. Therefore, NSGA-II is employed in this work to solve the MOO problems associated with the SMR process. In MOO, especially when dealing with conflicting objectives, metaheuristic algorithms (e.g., NSGA-II) are commonly used to generate a diverse set of Pareto-optimal solutions, also referred to as non-dominated solutions, Pareto-optimal front, or Pareto frontier (Wang & Rangaiah, 2017). These solutions are considered equally optimal with respect to the defined objectives (Wang et al., 2020), as each represents a trade-off scenario in which the improvement of one objective results in the deterioration of at least one other objective. Because the underlying search operates on a population rather than a single incumbent solution, metaheuristic MOO can explore multiple regions of a highly non-linear design space in parallel, reducing the risk of premature convergence to local optima (Talbi et al. 2012).

However, in real-world decision-making, selecting a single and actionable solution from this set of Pareto-optimal solutions is essential (Wang et al., 2023). Multi-criteria decision-making (MCDM) methods can be coupled with MOO to rank and recommend a preferred solution based on user-defined preferences or performance criteria (Wu et al., 2025). A comprehensive review by Rangaiah et al. (2020) noted that many MOO applications in chemical engineering have traditionally focused on generating the Pareto-optimal front, while comparatively less attention has been paid to the subsequent decision-making step using MCDM methods. This oversight may stem from the assumption that final solution selection can be adequately handled through engineering intuition and expert judgment. However, recent studies emphasize that relying solely on subjective preference can overlook structured, reproducible, and more optimal decision-making strategies that MCDM methods offer (Kesireddy et al., 2020; Ferdous et al., 2024).

Generally, MCDM methods provide a structured approach to evaluate and rank a set of alternatives (i.e., solutions) based on multiple, often conflicting, criteria (i.e., objectives), facilitating informed and rational decision-making in complex problems (Baydaş et al., 2024; Stević et al., 2024). Yang et al. (2023) employed an MCDM framework to evaluate low carbon fuel technologies across 13 criteria under economic, environmental, technical, and social dimensions. Park et al. (2023) applied MCDM methods to select the most preferred solution from the NSGA-II-generated Pareto-optimal front for a zero energy building system. Additionally, Nabavi et al. (2023, 2025) conducted comprehensive and systematic sensitivity analyses of multiple MCDM methods across various engineering optimization applications. Wang et al. (2024) provided a thorough review of the use of MCDM methods in chemical and process engineering, systematically detailing the key steps of the decision-making process, including normalization techniques, weighting schemes, and various MCDM approaches essential for decision-making. In the present study, two MCDM methods: technique for order of preference by similarity to ideal solution (TOPSIS) proposed in Hwang & Yoon (1981) and simplified preference ranking on the basis of ideal-average distance (sPROBID) introduced in Wang et al. (2021) are employed to rank the Pareto-optimal solutions and ultimately recommend one of them for implementation.

The rest of this paper is organized as follows. Section 2 describes the development of a hybrid surrogate model for the SMR reactor using artificial neural networks (ANN) trained on simulation data generated by a detailed one-dimensional mathematical model. Section 3 outlines the MOO problem formulation and introduces the two MCDM methods (i.e., TOPSIS and sPROBID). Section 4 presents the simulation results, including a detailed sensitivity analysis of the decision variables, followed by three optimization scenarios: maximizing methane conversion and hydrogen production; maximizing hydrogen output while minimizing carbon dioxide emissions; and a comprehensive three-objective case combining all goals. The optimization results and selection of preferred solutions using MCDM techniques are also discussed. Finally, Section 5 concludes the current study and provides recommendations for future research..

## 2. Machine Learning-Assisted Surrogate Modeling

The hybrid model is developed to accelerate the prediction of species profiles for use in process optimization and control. To achieve this, the mathematical model is executed under various conditions to generate data required for developing an alternative model, such as an ANN. The ANN is then trained on the generated data, resulting in a hybrid model of the process. In this model, the neural network replaces the computationally intensive part of the mathematical model. This substitution is expected to significantly speed up the calculations and enable faster prediction of species profiles. The reactor simulation is performed using the hybrid model, and finally, the accuracy of the neural network-based hybrid model is evaluated and compared.

To generate and collect the required data for model building, the Taguchi design of experiments method was used. This method offers the advantage of significantly reducing the number of tests needed to determine the relationships between input and output parameters. The experiment was designed using the Taguchi method for different levels of decision variables in the optimization process, as shown in Table 1. The various scenarios generated using the Taguchi design were individually simulated in the mathematical model, and the corresponding input and output data were recorded for training the neural network. The neural network inputs consisted of the temperature and partial pressures of the species during the reactor process, while the reaction rates served as the network outputs. A total of 1600 patterns were obtained using the Taguchi method, with all patterns initially generated randomly. To achieve more accurate results, the input and output patterns were normalized to a range of [-1, 1]. The detailed reactor model equations are provided in the Supporting Information.

Table 1. Data matrix of Taguchi method to generate data

| Run | $T_{in}$ | $F_{in}$ | St/CH$_4$ | H$_2$/CO | CO$_2$/CH$_4$ |
|-----|------|------|--------|--------|---------|
| 1 | 700 | 500 | 1 | 0.01 | 1 |
| 2 | 700 | 500 | 1 | 0.15 | 1.33 |
| 3 | 900 | 800 | 1 | 0.3 | 1.66 |
| 4 | 900 | 800 | 1 | 0.5 | 2 |
| 5 | 900 | 800 | 2 | 0.15 | 1 |
| 6 | 900 | 800 | 2 | 0.01 | 1.33 |
| 7 | 700 | 500 | 2 | 0.5 | 1.66 |
| 8 | 700 | 500 | 2 | 0.3 | 2 |
| 9 | 700 | 800 | 3 | 0.3 | 1 |

| 10 | 700 | 800 | 3 | 0.5 | 1.33 |
|----|-----|-----|---|------|------|
| 11 | 900 | 500 | 3 | 0.01 | 1.66 |
| 12 | 900 | 500 | 3 | 0.15 | 2 |
| 13 | 900 | 500 | 4 | 0.5 | 1 |
| 14 | 900 | 500 | 4 | 0.3 | 1.33 |
| 15 | 700 | 800 | 4 | 0.15 | 1.66 |
| 16 | 700 | 800 | 4 | 0.01 | 2 |

Denormalized outputs were used to calculate network errors. To build the neural network model, the input and output patterns were randomly divided into three groups with different proportions for training and testing purposes. Specifically, 60% of the patterns were used for training the network, 20% for validation, and the remaining 20% for testing the model, as shown in Figure 1.

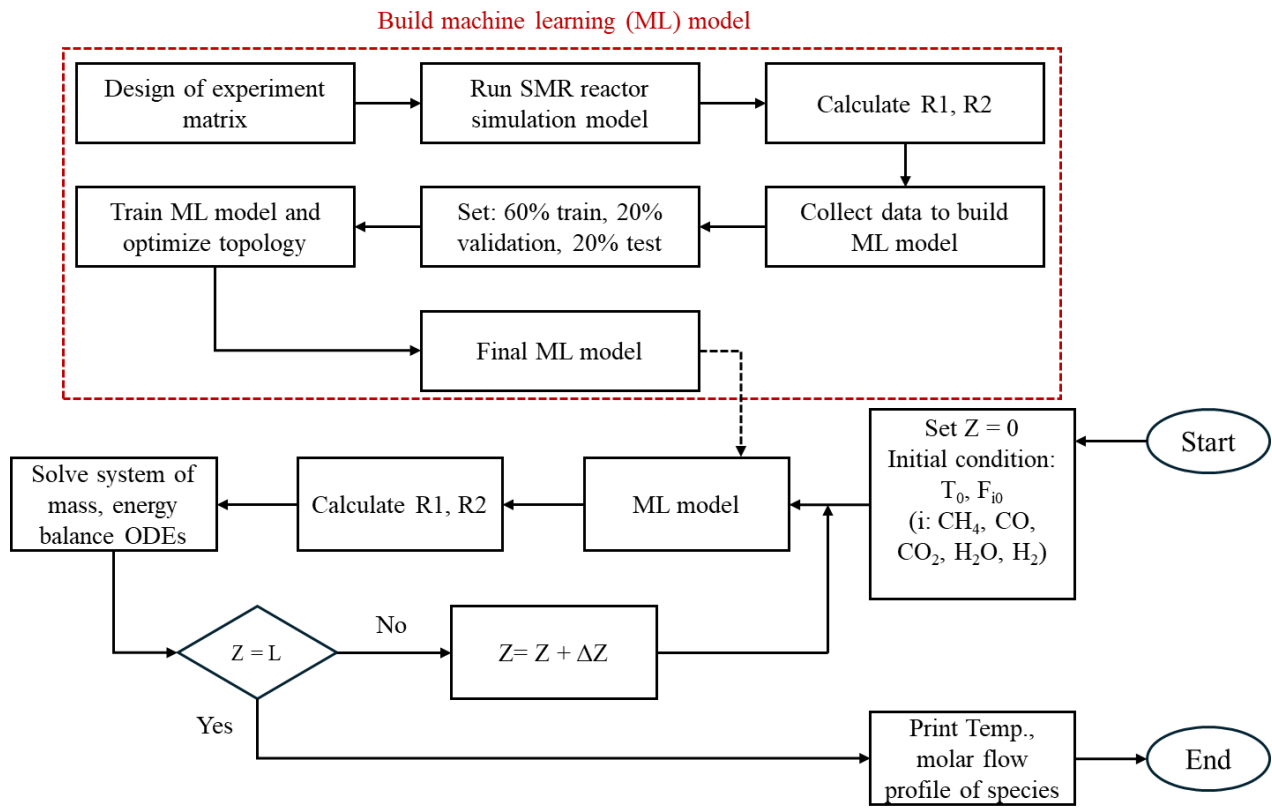

Figure 1. Flow chart of building surrogate model using machine learning tool for SMR reactor simulation

## 3. MOO and MCDM Procedures

### 3.1. MOO Formulation

MOO involves adjusting the decision or operational variables with the goal of achieving multiple optimization objectives simultaneously. A general mathematical representation of a MOO problem is as follows:

$$\text{Minimize } \mathbf{F}(\vec{X}) = \left[ f_1(\vec{X}), f_2(\vec{X}), \dots, f_k(\vec{X}) \right]^T$$

Subject to:

$$\vec{g}(\vec{X}) \leq 0$$

$$\vec{h}(\vec{X}) = 0$$

where $\vec{X} \in \mathbb{R}^n$ represents the decision variables, $\mathbf{F}(\vec{X}) \in \mathbb{R}^k$ the objective functions, $\vec{g}(\vec{X}) \in \mathbb{R}^m$ the inequality constraints, and $\vec{h}(\vec{X}) \in \mathbb{R}^q$ the equality constraints. Here, $k$, $n$, $m$, and $q$ denote the number of objective functions, decision variables, inequality constraints, and equality constraints, respectively. In contrast to single-objective optimization, MOO typically results in a collection of Pareto-optimal solutions, where no single solution is superior across all objectives. In this present research for MOO of the SMR process using the hybrid model, its decision variables (i.e., inlet temperature, feed flow rate, steam-to-methane ratio (St/CH$_4$), hydrogen-to-carbon monoxide ratio (H$_2$/CO), carbon dioxide-to-methane ratio, CO$_2$/CH$_4$), along with their operating lower and upper bounds, are presented in Table 2.

Table 2. The lower and upper bounds of decision variables for SMR process

| Decision variable | Lower bound | Upper bound |
| --- | --- | --- |
| T$_{in}$ (K) | 700 | 900 |
| F$_{in}$ (mol/s) | 500 | 800 |
| St/CH$_4$ | 1 | 4 |
| H$_2$/CO | 0.01 | 0.5 |
| CO$_2$/CH$_4$ | 1 | 2 |

The range of each variable was selected with reports available in available sources and available industrial data (Rajesh et al., 2000; Shahkarami & Fatemi, 2015). In addition, certain physical limitations of the SMR reactor have been taken into account. Specifically, the wall temperature of the reformer tubes is constrained based on the mechanical strength of the alloy used in their construction, as well as safety considerations aimed at preventing deformation or structural failure. The lower limit of the inlet gas temperature is determined according to the thermodynamic limitations and reaction rate, and the upper limit is determined according to the maximum heating of the feed by the fuel gases in the displacement part. The lower limit of the input $St/CH_4$ is to prevent the formation of coke on the catalyst, and in amounts lower than this limit, there is a possibility of the formation of coke on the catalyst. The high values of this parameter have an adverse effect on the economy of the process because this amount of excess steam takes a lot of heat to reach the outlet temperature in the reactor and cools down after the reformer. The lower limit of the input $H_2/CH_4$ is based on the minimum amount of hydrogen necessary to prevent the catalyst from becoming inactive at the reactor inlet, and its upper limit is to prevent the unnecessary return of hydrogen.

Three MOO scenarios are investigated in this study, namely, the simultaneous maximization of methane conversion and hydrogen yield; the maximization of hydrogen yield and minimization of carbon dioxide emissions; and the joint maximization of methane conversion and hydrogen yield along with the minimization of carbon dioxide emissions. NSGA-II is employed to solve these MOO problems and generate the corresponding Pareto frontiers.

### 3.2. MCDM Methods

### 3.2.1 Technique for order of preference by similarity to ideal solution (TOPSIS)

Originally introduced in Hwang & Yoon (1981) and widely used in many applications (Pandey et al., 2023), the TOPSIS method ranks alternatives based on their relative distances to the ideal solutions, with the top-ranked alternative being the one closest to the positive-ideal solution (i.e., ideal best) and farthest from the negative-ideal solution (i.e., ideal worst). The brief steps of this method are as follows. See Hwang & Yoon (1981) for its details.

**Step 1:** Normalize the objective matrix $f_{ij}$, with $m$ rows (solutions or alternatives) and $n$ columns (objectives or criteria) using vector normalization:

$$F_{ij} = \frac{f_{ij}}{\sqrt{\sum_{i=1}^{m} F_{ij}^2}}$$

**Step 2:** Generate the weighted normalized matrix $V_{ij}$ by multiplying each criterion by its given weight $w_j$:

$$V_{ij} = w_j F_{ij}$$

**Step 3:** Identify the ideal best $V^+$ and ideal worst $V^-$ solutions:

$$V_j^+ = \begin{cases} \max_i V_{ij}, & j \in J^+ \\ \min_i V_{ij}, & j \in J^- \end{cases}, \quad V_j^- = \begin{cases} \min_i V_{ij}, & j \in J^+ \\ \max_i V_{ij}, & j \in J^- \end{cases}$$

where $J^+$ and $J^-$ represent the sets of maximization and minimization objectives, respectively.

**Step 4:** Compute the Euclidean distances of each solution from the ideal best and ideal worst reference solutions:

$$S_i^+ = \sqrt{\sum_{j=1}^{n} \left(V_{ij} - V_j^+\right)^2}, \quad S_i^- = \sqrt{\sum_{j=1}^{n} \left(V_{ij} - V_j^-\right)^2}$$

**Step 5:** Calculate the relative closeness $C_i$ of each alternative as follows.

$$C_i = \frac{S_i^-}{S_i^+ + S_i^-}$$

In the end, the solution with the highest $C_i$ is top-ranked and recommended to decision-makers.

**3.2.2. Simplified preference ranking on the basis of ideal-average distance (sPROBID)**

The original PROBID method, proposed by Wang et al. (2021), distinguishes itself by incorporating multiple tiers of reference solutions when ranking non-dominated alternatives. These reference solutions span from the most favorable (that is, positive ideal solution; or in short: PIS) through successive ranks (i.e., 2nd PIS, 3rd PIS, and so forth) down to the least favorable, known as the negative ideal solution (NIS). The sPROBID method is a simplified variant of the original more sophisticated PROBID method. The steps of sPROBID are as follows:

**Step 1.** Normalize objective matrix with $m$ rows (solutions or alternatives) and $n$ columns (objectives or criteria) by using vector normalization.

$$F_{ij} = \frac{f_{ij}}{\sqrt{\sum_{k=1}^{m} f_{kj}^2}}$$

**Step 2.** Construct the weighted normalized objective matrix by multiplying each column with its assigned weight, $w_j$:

$$v_{ij} = F_{ij} \times w_j$$

**Step 3.** Determine the most PIS ($A_{(1)}$), 2nd PIS ($A_{(2)}$), 3rd PIS ($A_{(3)}$), …, and $m$th PIS ($A_{(m)}$) (i.e., the most NIS).

$$A_{(k)} = \left\{ \left( Large(v_j, k) \middle| j \in J \right), \qquad \left( Small(v_j, k) \middle| j \in J' \right) \right\}$$

$$= \left\{ v_{(k)1}, v_{(k)2}, v_{(k)3}, \dots, v_{(k)j}, \dots, v_{(k)n} \right\}$$

Here, $k \in \{1, 2, \dots, m\}$, $J$ = set of maximization/benefit objectives from $\{1, 2, 3, 4, \dots, n\}$, $J'$ = set of minimization/cost objectives from $\{1, 2, 3, 4, \dots, n\}$, $Large(v_j, k)$ means the $k$th largest value in the $j$th weighted normalized objective column (i.e., $v_j$) and $Small(v_j, k)$ means the $k$th smallest value in the $j$th weighted normalized objective column (i.e., $v_j$).

**Step 4.** Calculate the Euclidean distance of each solution to each of the $m$ ideal solutions:

$$S_{i(k)} = \sqrt{\sum_{j=1}^{n} (v_{ij} - v_{(k)j})^2}$$

**Step 5.** Here, sPROBID considers only the top and bottom quarters of ideal solutions for finding $S_{i(pos-ideal)}$ and $S_{i(neg-ideal)}$, respectively.

$$S_{i(pos-ideal)} = \begin{cases} \sum_{k=1}^{m\backslash 4} \dfrac{1}{k} S_{i(k)} & \text{when } m \geq 4 \\ S_{i(1)} & \text{when } 0 < m < 4 \end{cases}$$

Here, $m\backslash 4$ is the integer quotient of m divided by 4, which discards the remainder and retains only the integer portion.

$$S_{i(neg-ideal)} = \begin{cases} \sum_{k=m+1-(m\backslash 4)}^{m} \dfrac{1}{m-k+1} S_{i(k)} & \text{when } m \geq 4 \\ S_{i(m)} & \text{when } 0 < m < 4 \end{cases}$$

Here, $m + 1 - (m\backslash 4)$ gives the starting position of calculating negative-ideal distance.

**Step 6.** For sPROBID, the performance score is simplified to the ratio of negative-ideal distance over positive-ideal distance.

$$P_i = \frac{S_{i(neg-ideal)}}{S_{i(pos-ideal)}}$$

The farther a solution is from the group of negative ideal solutions and the closer it is to the group of positive ideal solutions, the higher the performance score $P_i$. The solution with the highest $P_i$ is top-ranked and recommended to decision-makers. See the original work by Wang et al. (2021) for more details about the PROBID and sPROBID methods.

**4. Results and Discussion**

## 4.1. SMR Reactor Simulation

The simulation of the SMR catalytic reactor was performed using a one-dimensional fixed-bed reactor model, considering the internal mass transfer resistance within spherical catalyst particles. The mass, energy, and reaction kinetics equations (comprising both algebraic and differential forms) were solved simultaneously using MATLAB. From these solutions, temperature profiles, conversion rates, and species mole fractions along the reactor length were determined. The impact of varying key operating parameters on conversion efficiency and overall reactor performance was then analyzed. Detailed descriptions of the kinetic model, mass and energy balance equations are provided in the Supporting Information. The reactor was simulated under feed conditions including a flow rate of 5.4 mol/s, an inlet temperature of 793 K, and a total pressure of one atmosphere with a composition of 0.20, 0.37, 0.14, 0.13, and 0.16 for carbon monoxide, hydrogen, carbon dioxide, water and methane, respectively. The other simulation parameters are given in Table S1.

Refer to the Supporting Information for detailed simulation results and discussions on the effects of inlet temperature, $St/CH_4$ ratio and feed flow rate on reactor performance. In brief, the simulation results indicate that the process gas temperature gradually increases along the reactor, driven by the endothermic nature of the reactions and heat transfer from the reactor wall. However, the peak gas temperature remains below the wall temperature, adhering to thermal constraints associated with catalyst deactivation and mechanical stress. Methane and carbon dioxide both exhibit increasing conversion rates along the reactor length. Beyond approximately 200 cm, methane conversion becomes dominant due to the sharp temperature rise and the endothermic characteristics of the reaction. The $CO_2$ conversion decreases slightly beyond 600 cm due to second reaction. Raising the inlet temperature from 793 K to higher values such as 850 K and 900 K does not significantly increase conversion. Moreover, higher inlet temperatures require more fuel, leading to higher costs and greater greenhouse gas emissions. Therefore, increasing inlet temperature beyond a certain point is not recommended based on this simulation analysis. Increasing the St/CH$_4$ has a noticeable effect on both temperature and methane conversion profiles, particularly in the second half of the reactor. Higher ratios result in higher outlet temperatures and increased methane conversion. The H$_2$/CO output ratio also increases with higher steam ratios, which is beneficial for syngas quality. Increasing the feed flow rate decreases the residence time, leading to a reduction in methane conversion along the reactor. The temperature profile also shows a plateau in the latter part of the reactor at higher flow rates. Despite lower conversion, the H$_2$/CO output ratio increases with flow rate due to greater overall hydrogen production.

## 4.2. Development of Neural Network Model

The neural network model is developed to approximate the thermodynamic reaction rate of the process, with the output layer comprising two neurons. Multiple networks with different topologies—varying in the number of neurons in the hidden layer—were trained and evaluated. Ultimately, the network with the lowest error on the validation data was selected as the optimal model. This optimal network contains 6 neurons in the input layer, 60 neurons in the hidden layer, and 2 neurons in the output layer. To achieve this, various configurations were tested, and the best-performing topology, illustrated in Figure 2, was chosen.

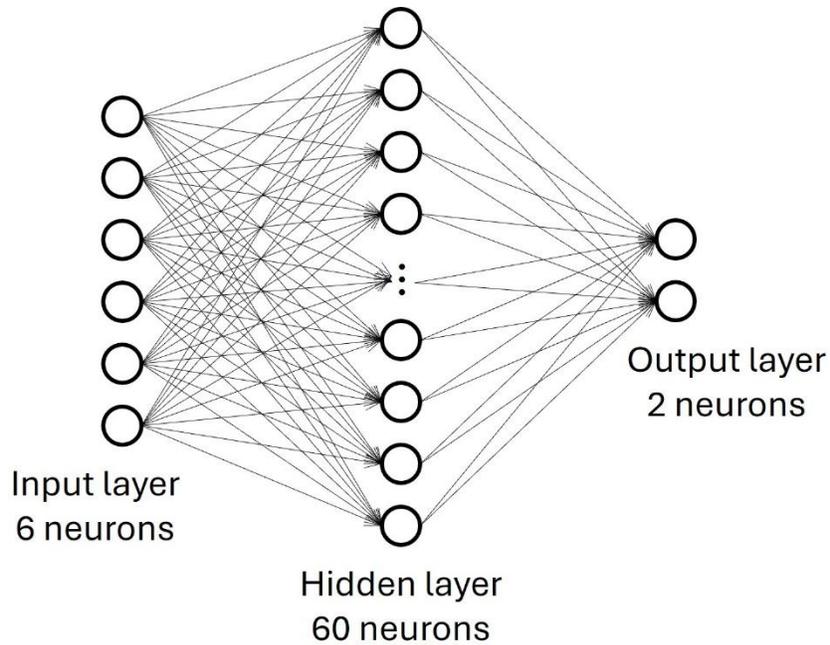

Figure 2. Optimal neural network topology with 6 neurons in the input, 60 neurons in the hidden layer and 2 neurons in the output layer

As shown in Figure 3, various network topologies were trained, and the mean squared error (MSE) values for training, validation, and test datasets are presented in a three-dimensional bar plot. The network configuration that produces the lowest overall MSE across all three datasets is deemed the most suitable for integration with the mathematical model. From the results, it can be observed that the topology with 60 neurons in the hidden layer provides the lowest total MSE, making it the optimal choice for network training and subsequent hybridization with the process model. The selected neural network consists of three layers: input, hidden and output. The activation functions used in these layers are as follows: tangent sigmoid (Tansig) in the hidden layer, and Purelin in the output layer. The network was trained using the Levenberg–Marquardt algorithm (Moré, 1978).

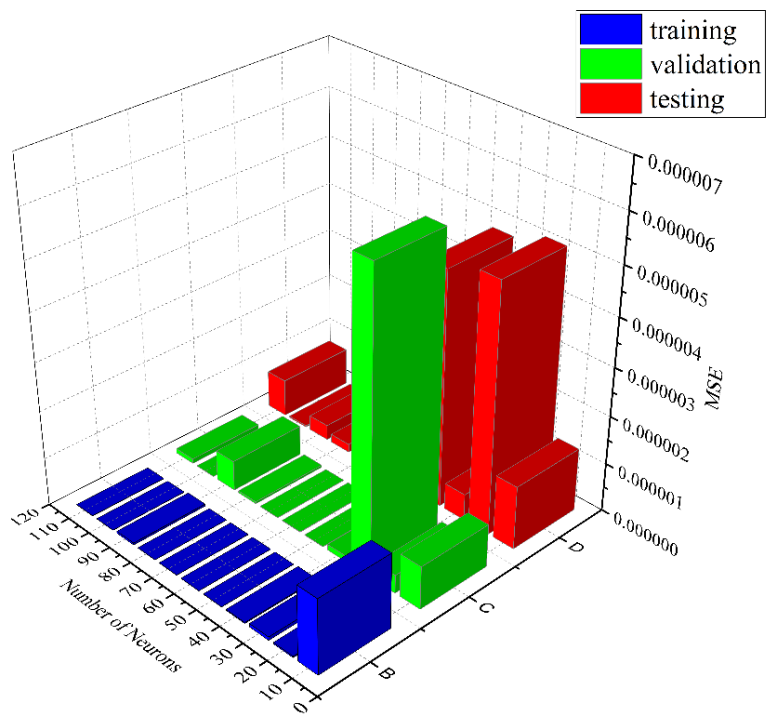

Figure 3. MSE values for training, validation, and test datasets across different neural network topologies.

To better evaluate and compare the neural network's performance in predicting reaction rates, the predicted results for the training, validation, and test datasets are plotted against the actual values in Figure 4. In these plots, the horizontal axis represents the target values from the mathematical model, while the vertical axis displays the values predicted by the neural network. The closer the data points are to the diagonal reference line and the less scattered they appear, the more accurately the network has captured the underlying relationship.

Additionally, the correlation coefficient between the predicted and actual values was calculated for each dataset. A value closer to 1 indicates a stronger predictive capability of the trained network. As seen in Figure 4, the predictions across training, validation, and test sets show a high degree of alignment with the reference line, indicating consistent weight assignment across the neurons and robust generalization performance. This demonstrates the strong ability of ANN to accurately estimate the reaction rates.

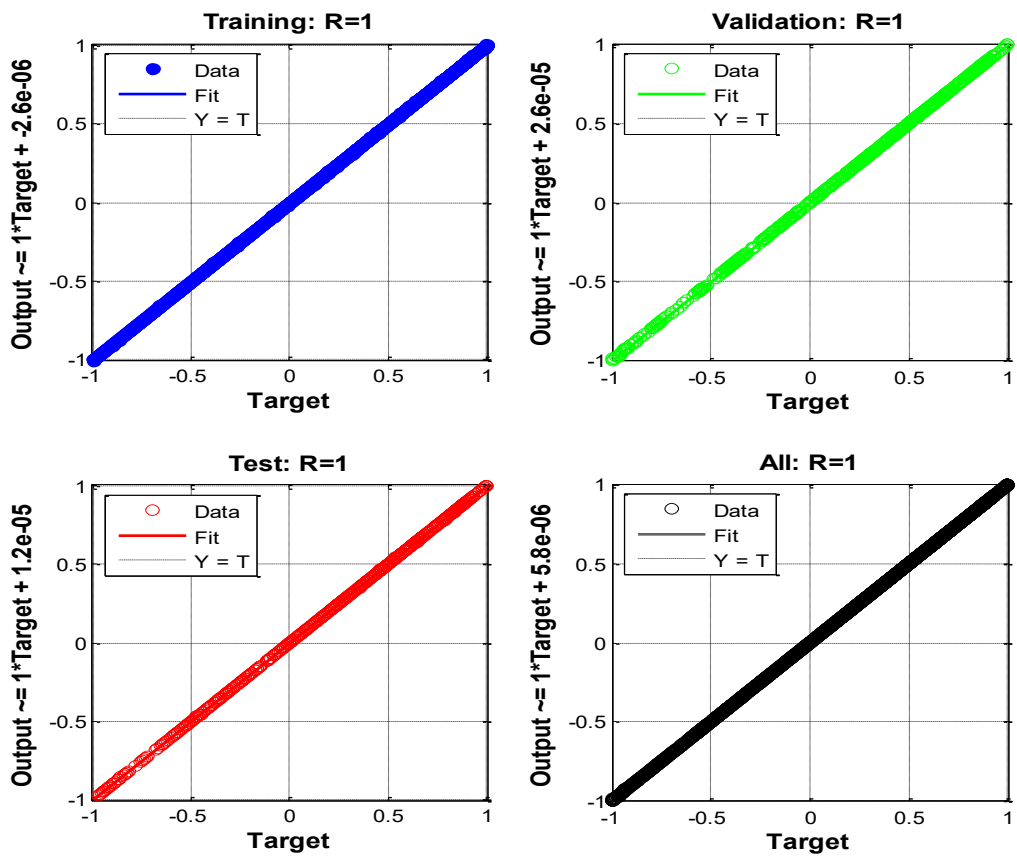

Figure 4. Neural network's prediction accuracy for reaction rates across training, validation, and test datasets.

## 4.3. Simulation of SMR Reactor using the ANN Hybrid Model

The reforming reactions occur within the catalyst grain and are often controlled by internal mass transfer limitations. A detailed one-dimensional model of this phenomenon involves the numerical solution of boundary value differential equations. Moreover, coupling these equations with the overall mass and energy balance equations significantly increases computational time, limiting the practicality of the mathematical model in process optimization tasks. To address this issue, a hybrid model is developed to accelerate the prediction of species profiles in the SMR reactor. The goal is to enhance computational efficiency and enable faster species profile predictions using the hybrid model. For this purpose, after training the neural network and replacing the detailed catalyst grain model with it, the hybrid model can estimate the overall reaction rate under any operating condition. Subsequently, the mass and energy balance equations are solved using this estimated rate.

Temperature and species profiles along the reactor can be obtained from the simulation results. Figure 5 shows the predicted reaction rates and the performance of the hybrid model compared to the mathematical model. As shown, the ANN predictions closely match those of the mathematical model. The temperature and mole fraction profiles calculated using the hybrid model align well with the results from the full mathematical simulation, demonstrating the accuracy of the hybrid approach.

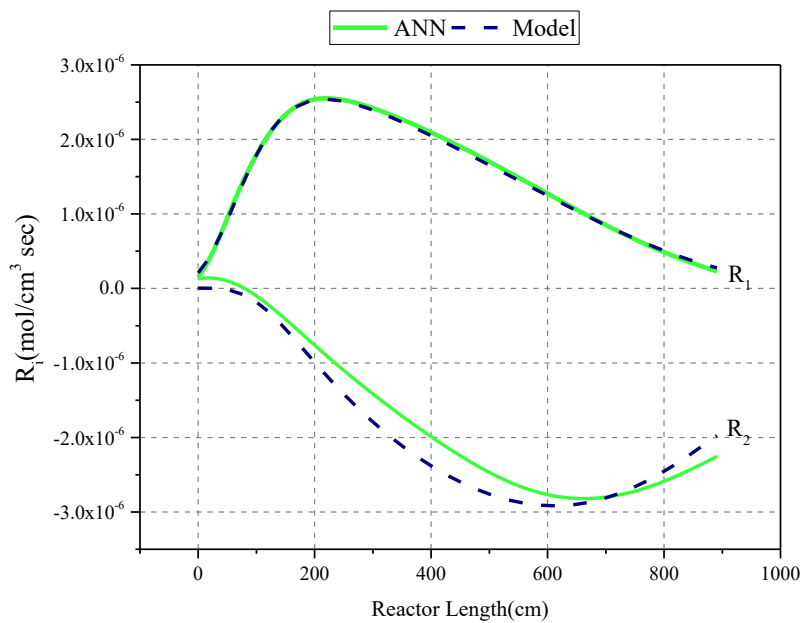

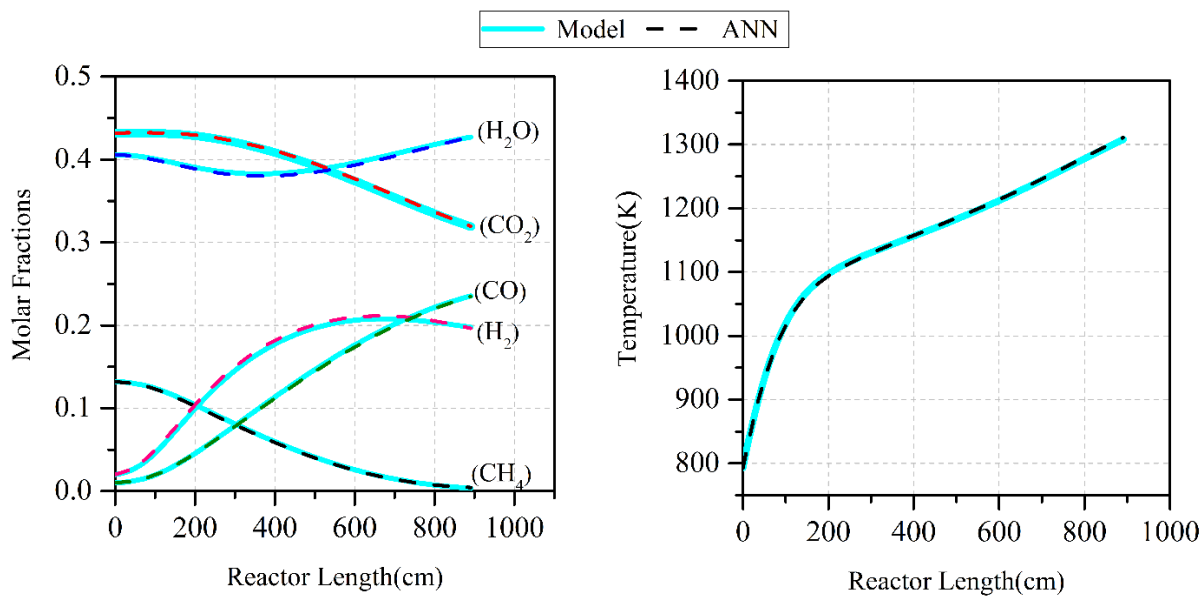

Figure 5. Comparisons of predicting the overall rate of reactions in the hybrid model with neural network and mathematical model

It is worth noting that the average simulation time for the SMR reactor was approximately 4 seconds using the neural network model, compared to 65 seconds with the mathematical model, as expected. In other words, the neural network model can save 93.8% of the simulation time compared with the mathematical model. Moreover, the results obtained from the neural network model demonstrated high accuracy and strong agreement with those of the mathematical model.

## 4.4. Sensitivity Analysis of Decision Variables

To identify the variables influencing the system, a sensitivity analysis was conducted on a set of selected parameters. In this approach, one variable was varied while keeping the others constant, and the resulting changes in each objective function were analyzed. Figure 6 presents the results of the sensitivity analysis of the objective functions with respect to the inlet temperature and $St/CH_4$ ratio. With the increase of inlet temperature, the methane conversion increases, while the production of carbon dioxide decreases slightly; Hydrogen production also increases slightly. Since the reforming reaction is endothermic, high temperature is favorable and convert more methane. Also, because the water-gas shift reaction is exothermic, high temperature suppresses this reaction. According to Le Châtelier's principle, with increasing temperature, endothermic reactions are progressed to the right side and exothermic reactions goes to the left side. Therefore, an increase in temperature causes more conversion of methane into carbon monoxide, carbon dioxide and hydrogen and less conversion of carbon monoxide into carbon dioxide. Obviously, the inlet temperature does not affect the methane feed. With the increase of $St/CH_4$, methane conversion increases. While the production of hydrogen and carbon dioxide is reduced. The reason behind this is the progress of reactions due to the presence of additional steam. Also, by increasing the $St/CH_4$ ratio, it is obvious that less methane is used in the feed.

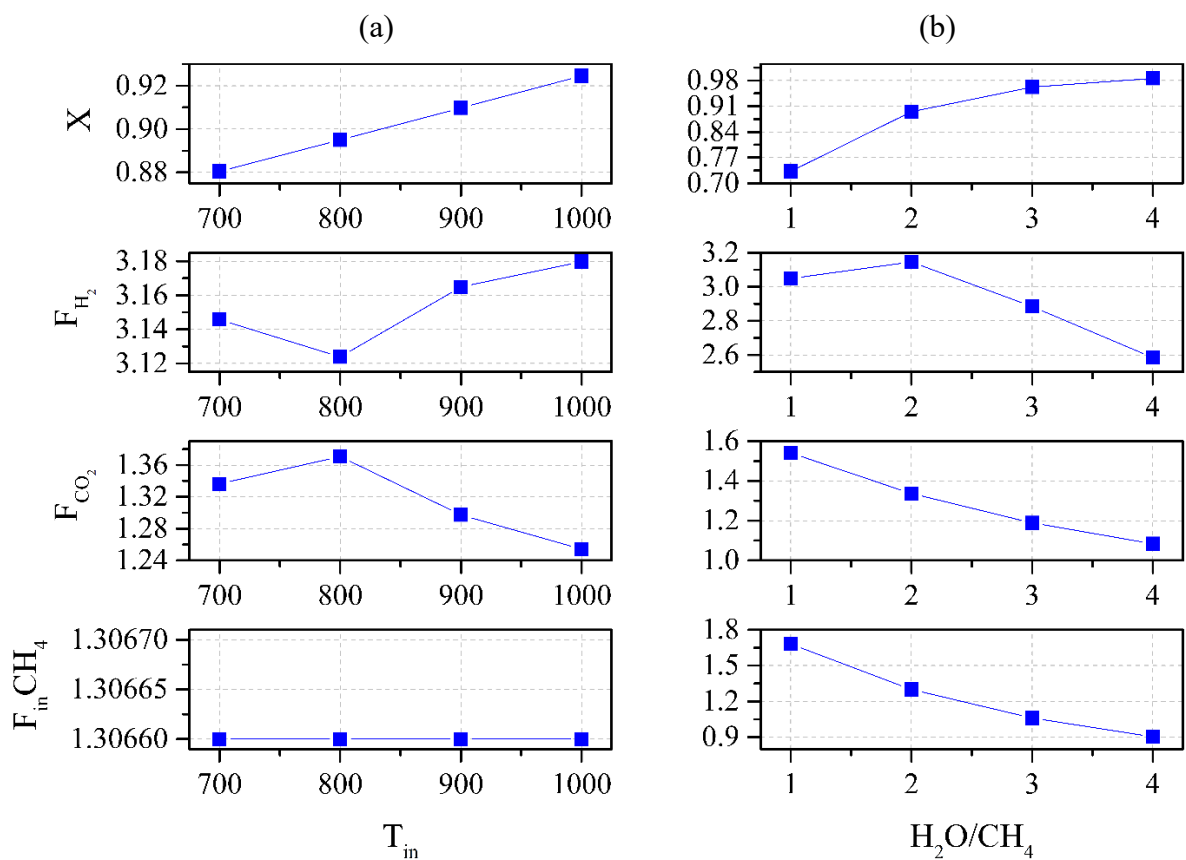

Figure 6. Sensitivity analysis of some selected objective functions to the inlet temperature (a) and steam to methane ratio (b)

Figure 7 shows that as the $CO_2/CH_4$ ratio increases, the methane conversion remains nearly constant, while hydrogen production decreases. Additionally, the carbon dioxide production increases, and the methane feed flow rate decreases accordingly.

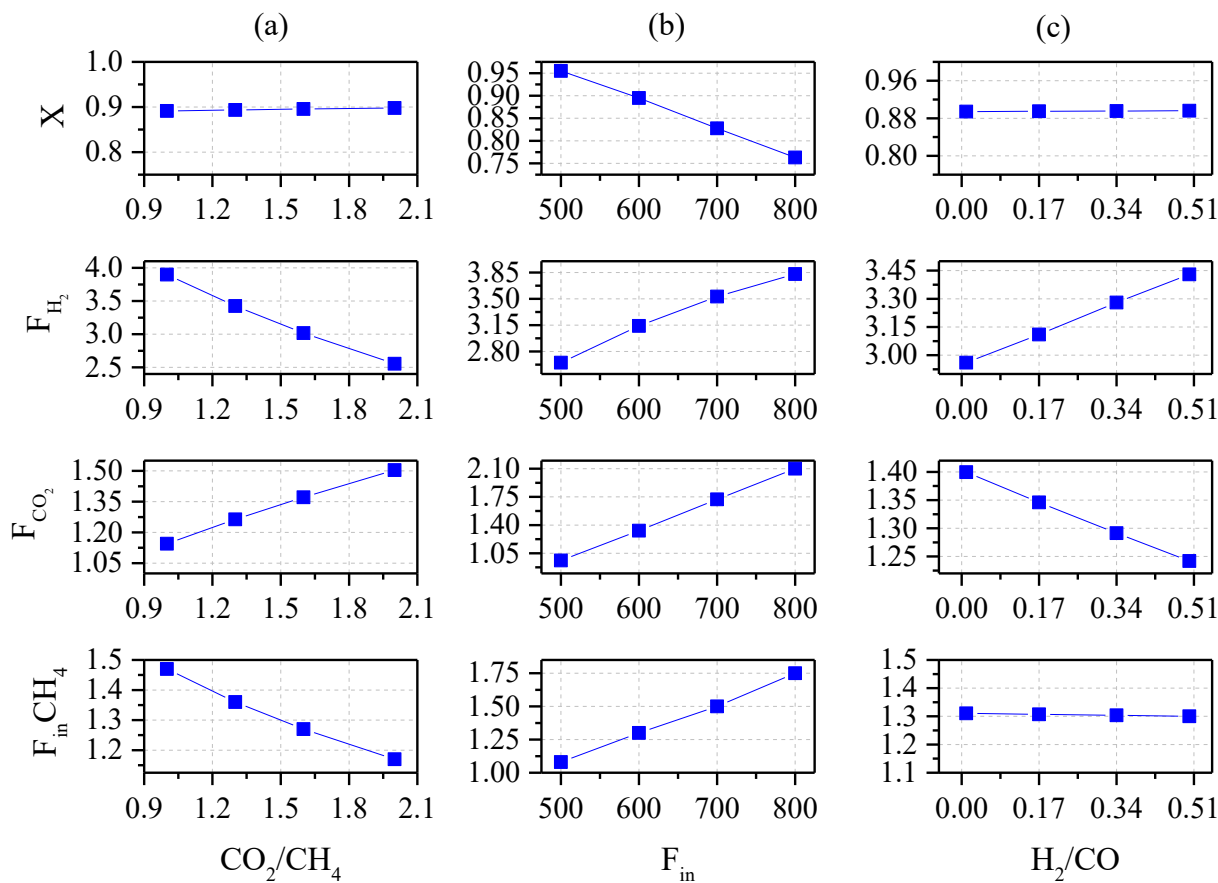

Figure 7. Sensitivity Analysis of the objective functions to the carbon dioxide to methane ratio, feed flow rate and hydrogen to carbon monoxide ratio

As the feed flow rate increases, the residence time within the reactor decreases, leading to a reduction in methane conversion. However, the overall hydrogen and carbon dioxide production increases indicating that a higher flow rate results in greater total product output, which is expected. In contrast, when the $H_2/CO$ increases, the methane conversion remains unchanged. Meanwhile, the production of carbon dioxide and monoxide decreases. However, hydrogen is produced more than the other.

## 4.4. Maximize Hydrogen Production and Methane Conversion (MOO Case I)

Figure 8 demonstrates the Pareto-optimal set obtained by simultaneously maximizing the methane conversion and the hydrogen output flow rate. As observed, moving along the curve from point A to point B results in an increase in methane conversion but a decrease in hydrogen output flow rate, confirming that the curve represents a Pareto-optimal front. An effective Pareto frontier should exhibit two key characteristics. First, it should span a sufficiently wide range for each objective function to provide the decision-maker with a variety of viable trade-off solutions. Second, it should demonstrate good diversity among the solutions, meaning the Pareto-optimal points are uniformly distributed across the front. As shown in Figure 8, these two characteristics are satisfied with the two objective functions considered.

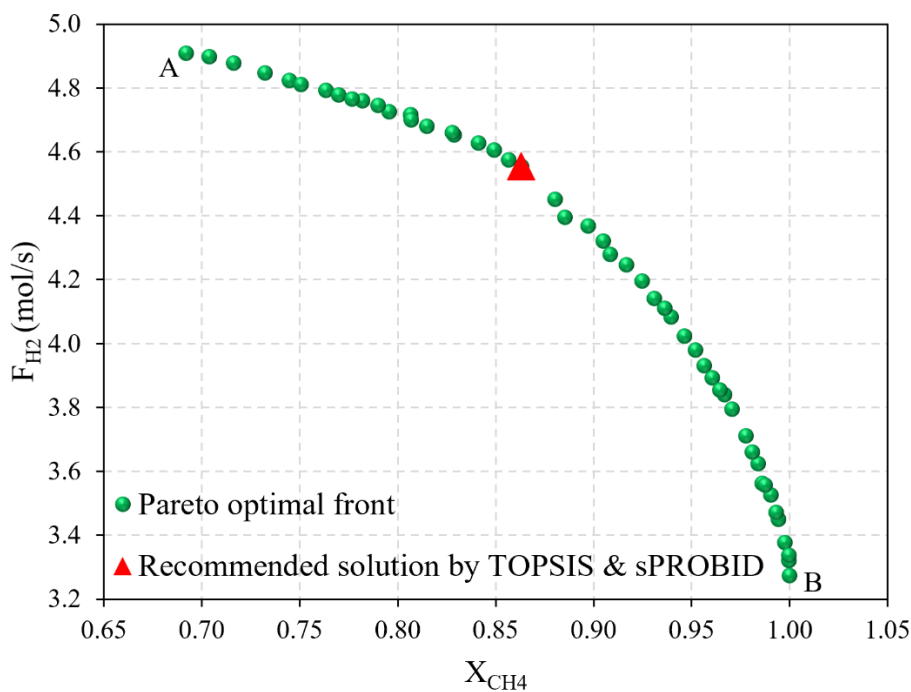

Figure 8. Pareto-optimal front for simultaneous maximization of methane conversion and output hydrogen flow rate, recommended solution by TOPSIS and sPROBID shown by ▲.

Figure 9 illustrates the convergence of the solution set toward the Pareto frontier over successive generations. The initial population is observed to be significantly distant from the true Pareto front.

However, as the number of generations increases, NSGA-II effectively evolves the population, progressively guiding the solutions closer to the optimal front.

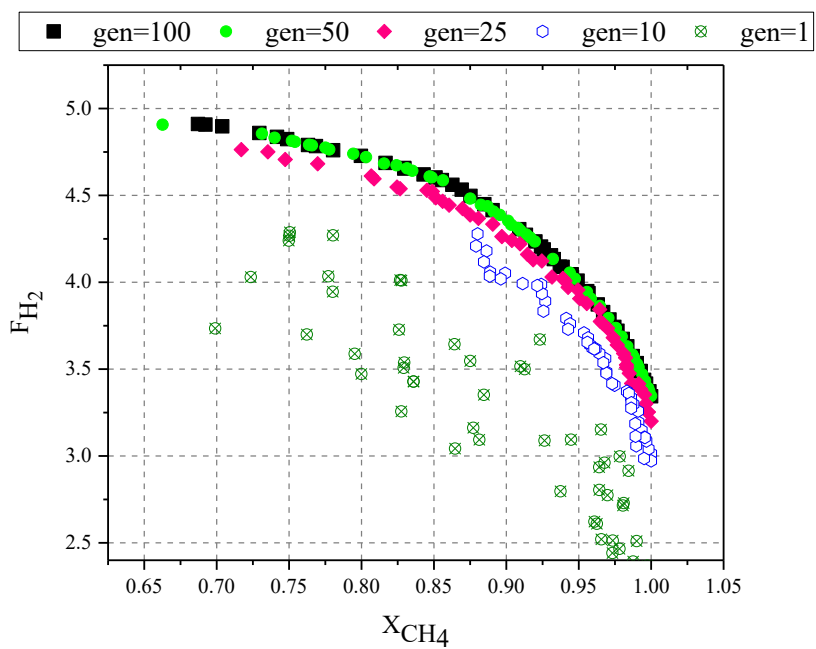

Figure 9. Pareto optimal front with different number of generations for simultaneous maximization of methane conversion rate and output hydrogen flow

The optimization times for the reactor using the mathematical model and the hybrid model were 8.24 hours and 2.9 hours, respectively. While the Pareto frontier illustrates the trade-off behavior among the objective functions, further insight can be gained by analyzing the behavior of the decision variables with respect to one of the objectives. This allows for understanding the contribution of each variable to the objective functions. Moreover, if the behavior of each variable aligns logically with the objectives, it confirms the reliability of the optimization results.

The influence of decision variables on the objective functions is depicted in Figure 10. As shown, the feed flow rate has the most significant influence on hydrogen production; increasing it results in a corresponding rise in hydrogen output. In response, the optimization algorithm progressively adjusts the feed flow rate toward its maximum value. Similarly, the optimal values for the inlet temperature and the $H_2/CO$ are pushed toward their upper limits to maximize hydrogen production.

In contrast, the optimal St/CH$_4$ suggests that higher values lead to reduced hydrogen output. Finally, the optimal CO$_2$/CH4 values are located at the lower end of the allowable range.

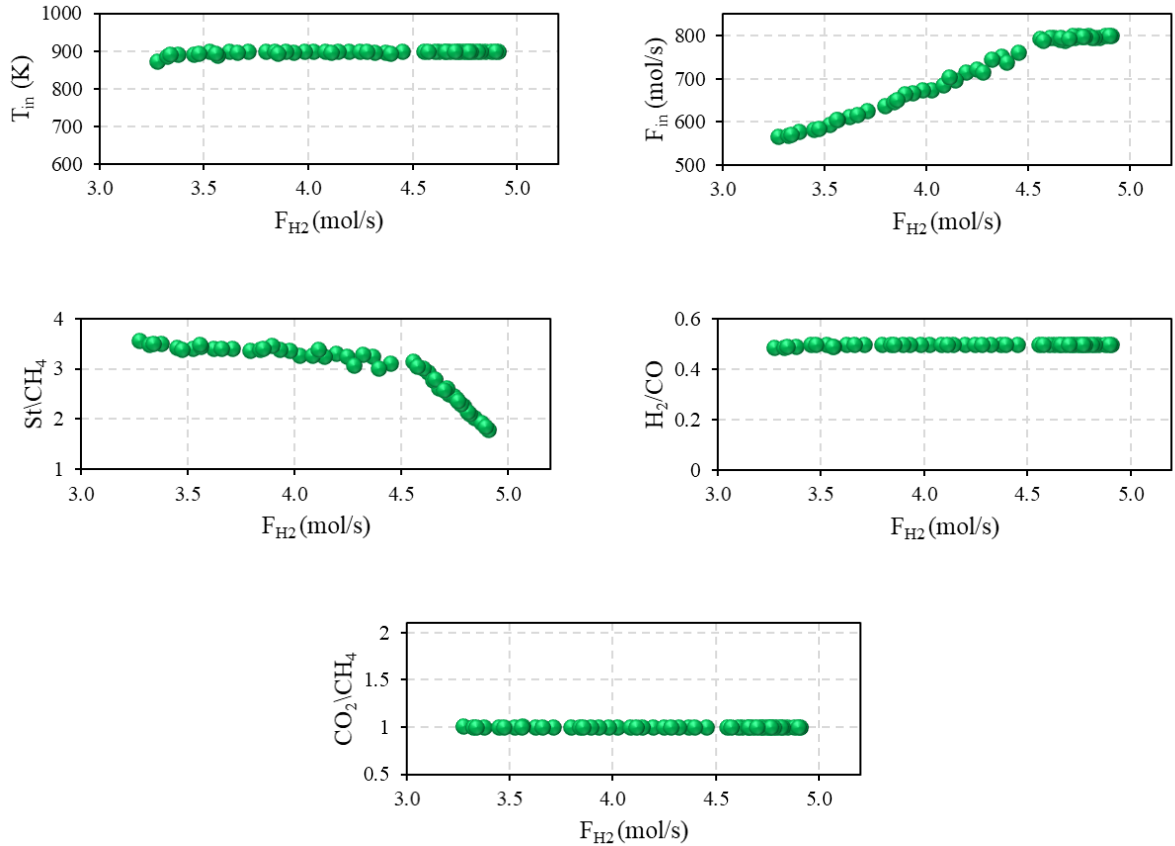

Figure 10. The effect of optimal values of decision variables on the hydrogen production (Case I)

Next, to select one solution from the Pareto-optimal front for practical implementation, both TOPSIS and sPROBID MCDM methods are employed. These methods rank all Pareto-optimal solutions and subsequently recommend the most favorable option based on computed ranking scores. In the present case study of simultaneous maximization of methane conversion and hydrogen output flow, equal weight (i.e., 0.5) is assigned to each objective in the MCDM process. The solution recommended by TOPSIS and sPROBID is indicated by the red triangle (▲) in Figure 8, corresponding to a methane conversion of 0.863 and a hydrogen production rate of 4.556 mol/s. For the decision variables at this selected operating point, the inlet temperature is at its upper limit of 900 K, while the feed flow rate is 793.2 mol/s, approaching its upper bound of 800 mol/s. The St/CH$_4$ is 3.14, the H$_2$/CO is fixed at its upper limit of 0.5, and the CO$_2$/CH$_4$ is at its lower bound of 1.

## 4.5. Minimize Carbon Dioxide Production and Maximize Output Hydrogen Flow (MOO Case II)

Figure 11 presents the Pareto-optimal set for the simultaneous minimization of carbon dioxide production and maximization of hydrogen output flow rate. As illustrated, an increase in hydrogen production is accompanied by a corresponding rise in carbon dioxide emissions, indicating a trade-off between the two objectives in this chemical process. This trade-off essentially stems from the underlying chemistry of the SMR process. Driving the reforming reaction to secure higher hydrogen production inevitably creates more carbon monoxide, which the downstream water-gas shift reaction then converts to carbon dioxide emissions. Because those two reactions favor opposite temperature directions, that is, steam reforming benefits from higher temperatures while the water-gas shift prefers cooler conditions, the optimal points naturally cluster near the upper end of the allowable temperature window, where hydrogen formation is vigorous, yet the shift reaction is partially suppressed. Similar qualitative trends have been reported for membrane-assisted or sorption-enhanced SMR units.

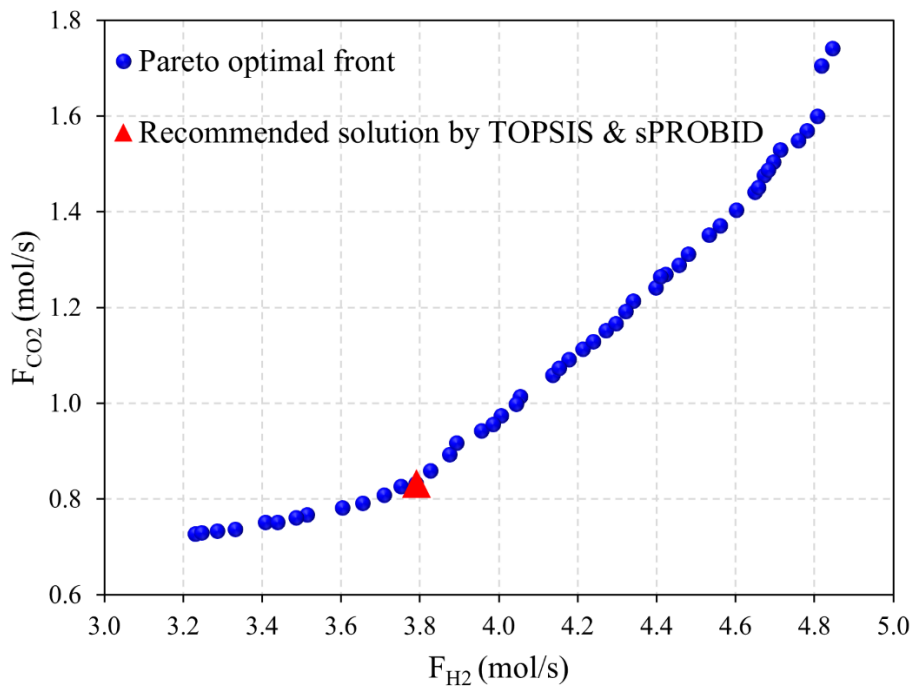

Figure 11. Pareto optimal set for minimizing carbon dioxide and maximizing hydrogen production, recommended solution by MCDM is shown by ▲

The behavior of the optimal values of the decision variables on minimizing carbon dioxide production is shown in Figure 12. Overall, feed flow rate dominates the objective: its scatter plot demonstrates an almost linear rise in carbon dioxide as feed flow rate increases. To curb carbon dioxide formation the algorithm therefore drives feed flow rate toward the lowest permissible values (i.e., 500 mol/s). Besides, the optimizer constantly holds the inlet temperature close to its upper bound of 900 K. For the St/$CH_4$, the curve is non-monotonic, but a higher ratio is generally preferred for lowering carbon dioxide production. The $H_2$/CO ratio is pinned at the upper limit of 0.5, and the $CO_2$/$CH_4$ ratio sits at the minimum feasible value of 1 across the entire design space, directly restricting the amount of carbon dioxide present in the feed and—by extension—in the product stream.

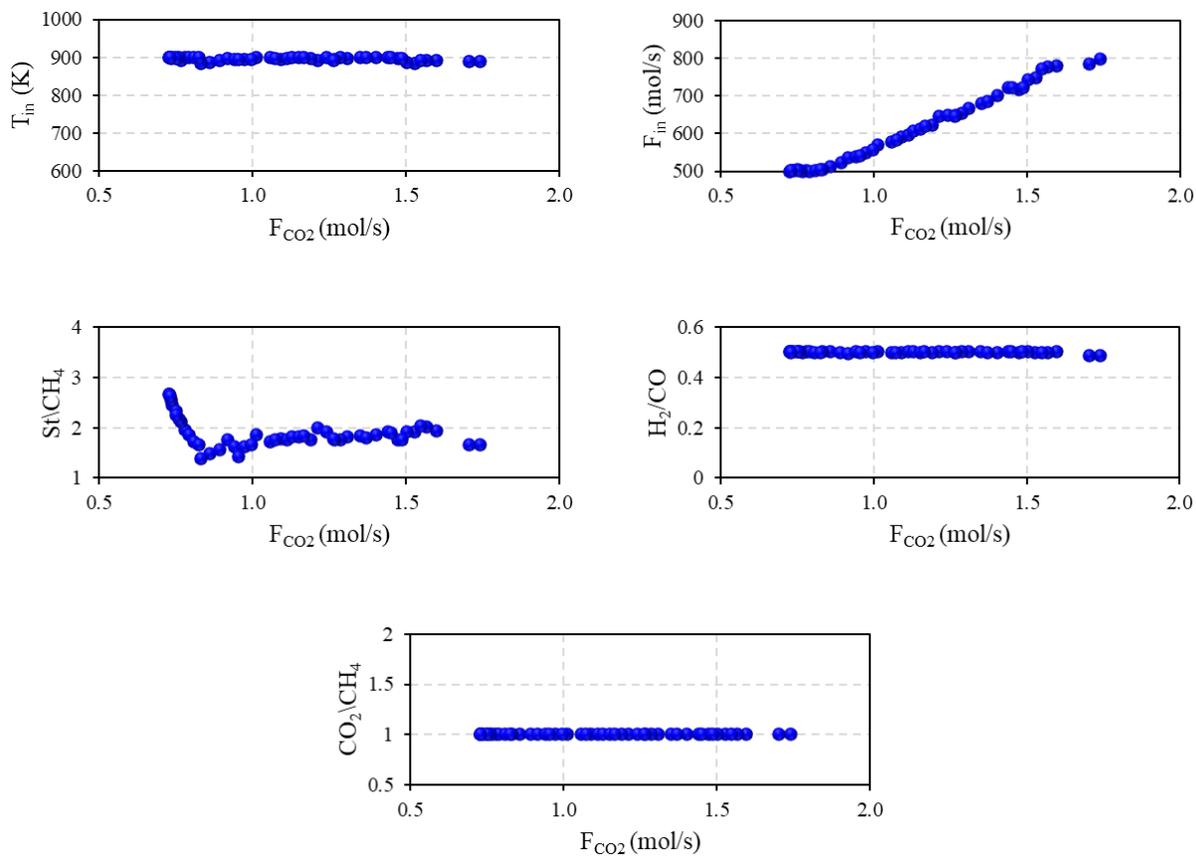

Figure 12. The effect of optimal values of decision variables on the minimization of carbon dioxide production

Subsequently, to select a single solution from the obtained Pareto-optimal front for practical implementation, both TOPSIS and sPROBID methods are again utilized to rank all Pareto-optimal solutions and recommend the most favorable one based on computed ranking scores. In this case study, which involves the simultaneous minimization of carbon dioxide production and maximization of hydrogen output flow, equal weighting (0.5) was assigned to each objective. The solution recommended by TOPSIS and sPROBID is represented by the red triangle ( ▲ ) in Figure 11, corresponding to a carbon dioxide production rate of 0.832 mol/s and a hydrogen output flow rate of 3.791 mol/s. At this selected operating point, the inlet temperature is set at 882.7 K, approaching its upper limit of 900 K. The feed flow rate is 503 mol/s, near its lower bound of 500 mol/s. The St/CH$_4$ is 1.38, the H$_2$/CO is fixed at its upper limit of 0.5, and the CO$_2$/CH$_4$ is maintained at its lower bound of 1.

## 4.6. Three-Objective Optimization: Maximize Methane Conversion and Hydrogen Output, and Minimize Carbon Dioxide Production (MOO Case III)

Based on the discussions in the preceding sections, it was observed that the two objective functions (i.e., maximizing both methane conversion and hydrogen output) exhibit a trade-off relationship. Similarly, a correlation exists between the production of hydrogen and carbon dioxide. Therefore, minimizing carbon dioxide production can be introduced as a third objective, alongside maximizing methane conversion and hydrogen output, in a three-objective optimization framework. Figure 13 presents the Pareto-optimal solutions for this three-objective problem. The shape of the resulting Pareto front indicates a broad range of trade-offs and a desirable level of solution diversity.

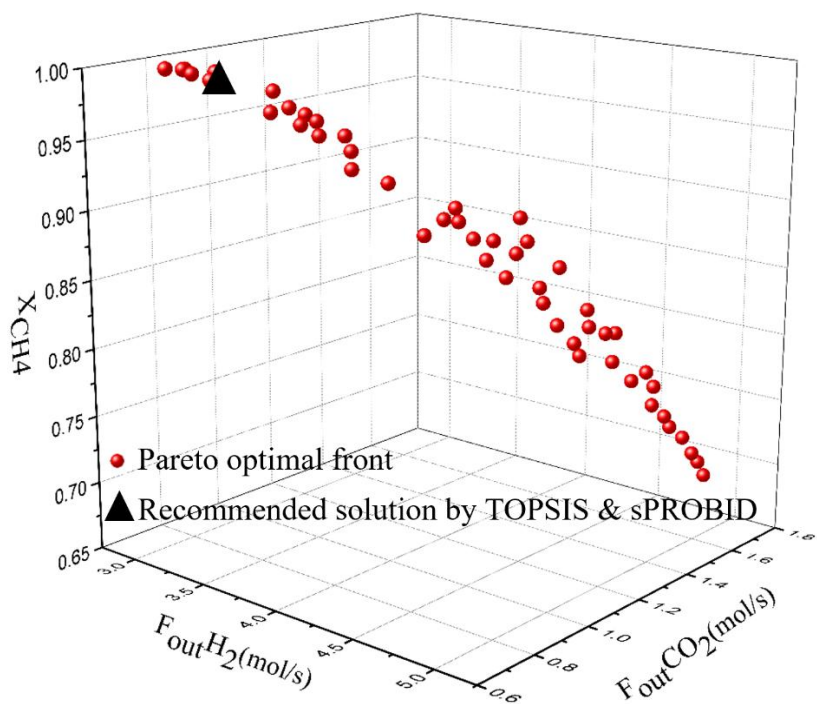

Figure 13. Pareto frontier for maximizing methane and hydrogen output conversion rate and minimizing emitted carbon dioxide

Figure 14 illustrates the influence of the optimal values of decision variables on the maximization of methane conversion. As shown, the inlet temperature is set at its upper limit. This is consistent across all three optimization objectives, highlighting the significant role of temperature in enhancing conversion. Elevated temperatures increase the reaction rate, thereby promoting higher methane conversion. Consequently, the optimization algorithm favors higher inlet temperatures to drive the reaction more effectively. The optimal feed flow rate values indicate that reducing the flow rate increases the residence time of methane within the reactor, which in turn enhances the conversion. This relationship is evident as the highest methane conversion is achieved when the feed flow rate approaches its lower limit. Regarding the $St/CH_4$, the results show a generally positive correlation with methane conversion: as the $St/CH_4$ increases, so does the conversion. Additionally, the $H_2/CO$ is fixed at its upper limit of 0.5, while the $CO_2/CH_4$ remains at its lower bound of 1.0 throughout the optimization.

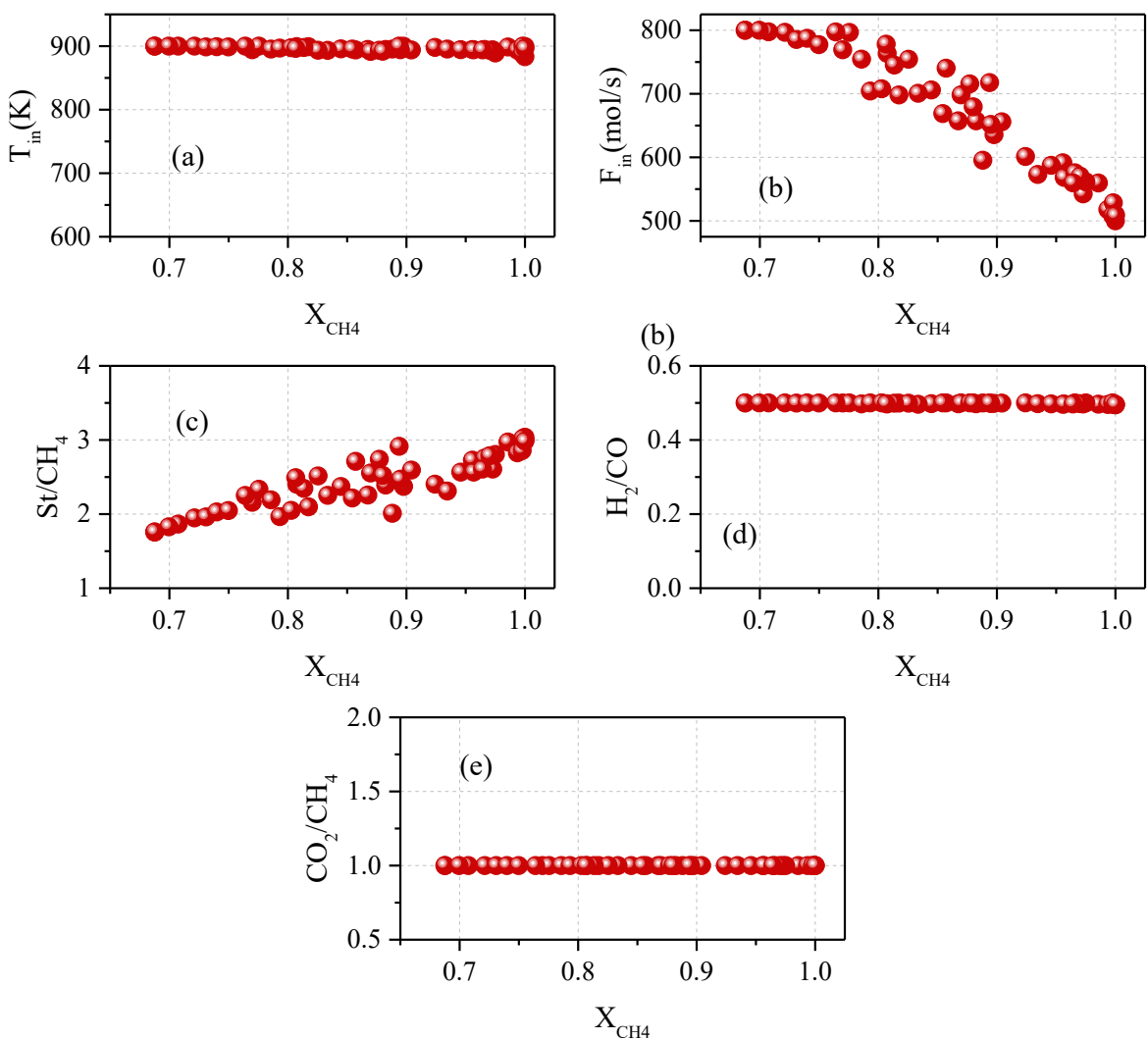

Figure 14. The effect of optimal values of decision variables on the objective function of maximizing methane conversion, Case III

Based on the discussion, it can be concluded that in the three-objective optimization of the SMR reactor, the objective functions (i.e., methane conversion, hydrogen output, and carbon dioxide production) exhibit inherent trade-off relationships. These trade-offs result in a set of Pareto-optimal solutions, within which the behavior of the decision variables is consistent with variations along the Pareto front. Furthermore, the obtained Pareto front spans a wide range, offering significant flexibility for users to select operationally favorable solutions. The TOPSIS and sPROBID methods are used to facilitate the selection of one solution from the Pareto-optimal front for ultimate implementation. Equal weighting (0.333) is assigned to each of the three objectives, namely, maximizing hydrogen output, minimizing carbon dioxide emissions, and maximizing methane conversion. The solution recommended by TOPSIS and sPROBID method is denoted by the black triangle (▲) in Figure 13, corresponding to a hydrogen output flow rate of 3.335 mol/s, a carbon dioxide production rate of 0.781 mol/s, and a methane conversion of 0.988. At this selected operating condition, the inlet temperature is supposed to be set at 885.8 K, approaching its upper limit. The feed flow rate is 520.7 mol/s. The St/$CH_4$ is 2.73, while the $H_2$/CO is fixed at its upper bound of 0.5, and the $CO_2$/$CH_4$ is maintained at its lower limit of 1.

## 5. Conclusion

In conclusion, this research illustrated a comprehensive modeling and optimization framework for a SMR reactor by integrating mechanistic simulation, machine learning-assisted surrogate modeling, and advanced MOO and MCDM methods. A detailed one-dimensional reactor model, which incorporates internal mass transfer resistance, was utilized to simulate reactor performance. To overcome the high computational cost associated with this detailed model, a hybrid model was constructed using an ANN. This ANN-based surrogate accurately predicted reactor behavior while reducing the average simulation time by 93.8% without compromising model accuracy. The hybrid model was then integrated into MOO for three different optimization cases: 1) maximizing methane conversion and hydrogen output; 2) maximizing hydrogen output while minimizing carbon dioxide emissions; and 3) a three-objective case combining all aforementioned goals. The NSGA-II method was employed to generate well-distributed Pareto-optimal front for each case. Additionally, two MCDM methods, namely, TOPSIS and sPROBID, were applied to select the optimal operating points from the Pareto-optimal fronts. In the first case study, the selected optimal solution achieved a methane conversion of 0.863 and a hydrogen output flow rate of 4.556 mol/s.

In the second case, the model achieved a hydrogen output of 3.791 mol/s while keeping carbon dioxide emissions at a low level of 0.832 mol/s. In the comprehensive three-objective scenario, the recommended solution yielded a hydrogen output of 3.335 mol/s, a carbon dioxide production of 0.781 mol/s, and an impressive methane conversion of 0.988. The results underscore the effectiveness of the hybrid modeling approach in significantly reducing computational demands while preserving high accuracy. Moreover, the systematic use of MOO and MCDM tools enabled rational selection of trade-off solutions tailored to specific process objectives. This framework provides a robust and generalizable methodology for optimizing catalytic reactor systems under multiple, often conflicting, objectives.